\documentclass[11pt]{article}      
\begin{document}      

\title{On a Linear Chaotic Quantum Harmonic Oscillator 
\footnote{Address for correspondence: Professor J. Duan,
         Department of Mathematical Sciences, Clemson University, 
         Clemson, South Carolina 29634, USA. E-mail: duan@math.clemson.edu; 
         Tel: (864)656-2730; Fax: (864)656-5230.   \protect  \\ 
 \indent        This work was supported by the USA National 
         Science Foundation Grant DMS-9704345 for J. Duan  and by
         the National Natural Science Foundation of China
         Grant No. 19572075 for X. Fu. }   } 

\author{     }

\date{September 7, 1997}

\maketitle

\begin{center}
 
 Jinqiao Duan$^{1}$ , Xin-Chu Fu$^{2,3}$, 
 Pei-De Liu$^{4}$, and Anthony Manning$^{2}$  \\ 

{\em 
1. Department of Mathematical Sciences, Clemson University,\\
   Clemson, South Carolina 29634, USA. \\  
2. Mathematics Institute, University of Warwick, \\
   Coventry CV4 7AL, England, UK.\\
3. Wuhan Institute of Physics and Mathematics, \\ 
   The Chinese Academy of Sciences,\\
   P. O. Box 71010, Wuhan 430071, CHINA.\\
4. Department of Mathematics, Wuhan University, \\ 
   Wuhan 430072, CHINA.\\ }  

\end{center}

\begin{abstract}
 
	We show that a linear quantum harmonic oscillator is chaotic
	in the sense of Li-Yorke.  We also prove that the weighted
	backward shift map, used as an infinite dimensional
	linear chaos model,
	in a separable Hilbert space is chaotic in the sense of 
	Li-Yorke, in addition to being chaotic in the sense of Devaney.
	

\end{abstract}

\section{Introduction}

We consider an unforced 
quantum harmonic oscillator, i.e., a very small frictionless mass-spring
system whose evolution is modeled by the Schr\"{o}dinger equation
(\cite{Liboff})
\begin{equation}
	i \hbar \psi_t = -\frac{\hbar^2}{2m} \psi_{xx} + \frac{k}{2} x^2\psi, 
\end{equation} 
with wave function $\psi(x,t)$, displacement $x$, mass $m$, stiffness $k$
and Planck number $\hbar$.
The nondimensionalized stationary states in the separable Hilbert space
$X=L^2(-\infty, \infty)$ form an orthonormal basis
\begin{equation}
	\psi_n(x) = e^{-x^2/2}H_n(x)/\sqrt{\sqrt{\pi} 2^n n!}, 
		n=0, 1,  \cdots,
\end{equation} 
where 
$$
	H_n(x) = (-1)^n e^{x^2}\frac{d^n}{dx^n} e^{-x^2},
$$
is the $n$th Hermite polynomial.
The natural space for the quantum harmonic oscillator is
the Schwartz class $F$ of rapidly decreasing functions in
$X=L^2(-\infty, \infty)$ as defined in section 2.
Gulisashvili and MacCluer (\cite{Gulisashvili_MacCluer})
defined a linear, closed, unbounded operator, namely, the weighted 
backward shift operator $B$ on the space F by
\begin{eqnarray} 
B: F & \rightarrow  & F,     \\
B\psi_n & \equiv  & \frac1{\sqrt{2}}(x+\frac{d}{dx})\psi_n  
= \sqrt{n} \psi_{n-1}.
\end{eqnarray}
$B$ has no resolvent set since every complex number
$\lambda$ is in the point spectrum of $B$. By using a result of Godefroy and 
Shapiro (\cite{Godefroy_Shapiro}),
Gulisashvili and MacCluer (\cite{Gulisashvili_MacCluer})
have shown that the shift operator $B$ is chaotic in  the sense of Devaney  
(\cite{Devaney}), namely, it has
topological transitivity (dense orbits), sensitivity to  
initial conditions (orbit divergence), and density of periodic points.

There is not a universally accepted definition of ``chaos".  Although the
definition  in Devaney \cite{Devaney} (or Wiggins \cite{Wiggins})
seems a popular one, other definitions
\cite{Kirchgraber}, which capture or describe other dynamical
behavior of a system, are proposed and used in modern nonlinear dynamics. 
Sometimes 
Li-Yorke's definition of chaos does characterize the complexity of dynamical 
systems. For example, for one-dimensional dynamical systems (the iteration of 
continuous self-maps on intervals), Li-Yorke's chaos is equivalent to having 
positive topological entropy. The same conclusion holds for subshifts of 
finite type. 

The relation between  various definitions of chaos is not always apparent, 
yet each definition
certainly describes important dynamical behavior, and there is a need to
understand different behavior in various systems.  This motivates us to study
chaos of the above weighted backward shift map $B$ in the sense of Li-Yorke,
in terms of asymptotic separation of orbits.

In this paper, we construct a chaotic set for the operator $B$, and therefore 
show that the above linear quantum harmonic 
oscillator is chaotic in the sense of Li-Yorke.

\section{Chaotic Set of Operator $B$}

We now show that the above weighted
shift operator $B$ is chaotic in the sense of Li-Yorke.
We first recall the chaos definition of Li-Yorke \cite{Li_Yorke}.

\noindent{\bf Definition }  Let $ M$ be a metric space with metric
$\rho$ and 
$f: M \rightarrow M $ be a continuous map. The discrete dynamical system 
$(M, f)$ is called chaotic in the sense of Li-Yorke if there exists an 
uncountable subset $S$ of nonwandering non-periodic points such that 
whenever $x, y \in S, x\neq y$ , the following conditions hold, 
$$
\left.\begin{array}{l} 
(i)\ \limsup\limits_{n\rightarrow +\infty } \rho (f^n(x), f^n(y))>0\\
(ii)\ \liminf\limits_{n\rightarrow +\infty } \rho (f^n(x),f^n(y))=0
\end{array}\right. ,  
$$
The subset $S$ above is called a chaotic set for $f$.

\it Remark \rm\ \ The original characterization of chaos in Li-Yorke's  
theorem \cite{Li_Yorke} is via three conditions. The third one is :
$$
(iii)\ \limsup\limits_{n\rightarrow +\infty } \rho(f^n (x),f^n(p))>0,\ \
\forall x\in S, \forall p\in P(f)
$$
This condition means that no point in $S$ is asymptotically periodic. 
From conditions (i) and (ii) in the Definition,  $S$ contains at 
most one asymptotically periodic point \cite{Zhou}. So condition (iii) is 
not essential and can be removed.

In the following, we  construct a chaotic set $S$ for the operator 
$B : F \rightarrow F, B \psi_n = \sqrt{n}\psi_{n-1}$.

In terms of the orthonormal basis $\{\psi_n\}$, 
the Schwartz class $F$ can be written as 
(\cite{Gulisashvili_MacCluer})
$$
F = \{ \phi \in L^2(-\infty, \infty): \phi = \sum_{n=0}^{\infty}c_n \psi_n, 
    \sum_{n=0}^{\infty}|c_n|^2 (n+1)^r<\infty, \forall r \geq 0\}.
$$
$F$ is an infinite-dimensional Fr\'{e}chet space with topology 
defined by the system of semi-norms $p_r(\cdot)$ of the form
(\cite{Yosida}) 
$$
p_r (\phi) = p_r (\sum_{n=0}^{\infty} c_n \psi_n) = 
(\sum_{n=0}^{\infty}|c_n|^2 (n+1)^r)^{1/2}, \ \ \ r\geq 0.
$$
This topology on $F$ is also given by the  metric $\rho$  
$$
\rho (\phi, \psi) = \sum_{m=0}^{\infty}2^{-m} p_m (\phi - \psi)\cdot 
             (1+p_m (\phi -\psi))^{-1}.
$$

Fix ${\theta} \in (0, 1)$ and define $\phi^{\theta} = 
\sum_{n=0}^{\infty}c_n^{\theta}\psi_n$  by
$$
\left\{\begin{array}{l} 
c^{\theta}_0=0 \\
c^{\theta}_n = \left\{\begin{array}{cl} 1/ \sqrt{n!}   & \mbox{ if $ 
n = k^2, [k{\theta}]-[(k-1){\theta}] = 1 $ }\\
0   & \mbox{otherwise ,} 
\end{array}\right. 
\end{array}\right.
$$ 
where $k=1, 2, \cdots$, and  $[\cdot]$  denotes the integer part of a real 
number.

Let  $S=\{\phi^{\theta}: {\theta}\in (0,1)\}$. From
$$
\sum_{n=0}^{\infty}|c_n^{\theta}|^2 (n+1)^r\leq 
\sum_{n=0}^{\infty}\frac{(n+1)^r}{n!}<\infty,  \ \ \ \forall r,
$$
we have $S \subseteq F$. 

Let $V(\phi^{\theta}, \varepsilon)$ denote  
$\{\phi \in F: \rho (\phi, \phi^{\theta})< \varepsilon\}$. Take
$$
\phi_N^{\theta} = \sum_{n=0}^{N} c_n^{\theta}\psi_n + 
\sum_{n=N+1}^{\infty}c_{n-(N+1)}^{\theta} \cdot 
(n(n-1) \cdots (n-N))^{-1/2}\psi_n.
$$
It is obvious that $\phi_N^{\theta} \in F$.  Moreover,
$\ \ \  \forall r \geq 0,$ we have
 
\begin{eqnarray} 
p_r(\phi_N^{\theta}-\phi^{\theta}) & = &
(\sum_{n=N+1}^{\infty}|c_n^{\theta} - c_{n-(N+1)}^{\theta} 
(n(n-1)\cdots(n-N))^{-1/2}|^2(n+1)^r)^{1/2}     \nonumber   \\ 
& \leq &   (\sum_{n=N+1}^{\infty}2(|c_n^{\theta}|^2+
|c_{n-(N+1)}^{\theta}|^2\cdot (n(n-1)\cdots(n-N)
)^{-1})(n+1)^r)^{1/2}    			 \nonumber \\ 
& \leq &   (\sum_{n=N+1}^{\infty}{4(n+1)^r}/{n!})^{1/2}.
\end{eqnarray}
This shows that $p_r(\phi_N^{\theta}-\phi^{\theta})\rightarrow 0 $
 as  $N\rightarrow \infty.$ Therefore, $\forall 
\varepsilon >0,$ there exists $N = N_{\varepsilon}$ big enough, such that 
$\phi_{N_{\varepsilon}}^{\theta} \in V(\phi^{\theta}, \varepsilon)$.  
Because 
$$
B^{N_{\varepsilon}+1}\phi_{N_{\varepsilon}}^{\theta} = \phi^{\theta},
$$
we know that
$$
(B^{N_{\varepsilon}+1} V (\phi^{\theta}, \varepsilon))\cap V (\phi^{\theta}, 
\varepsilon) \neq \emptyset .
$$
So all points in $S$ are nonwandering. It is obvious that all points in $S$
are nonperiodic as well.

Denote by $P(\phi^{\theta}, k)$ the number of $c_l^{\theta}$'s 
which satisfy $c_l^{\theta} \neq
0, 0\leq l\leq k.$ Then
$$
[\sqrt k {\theta}]\leq P(\phi^{\theta}, k)\leq [(\sqrt k +1)\theta],
$$
which implies that
$$
\lim_{k\rightarrow \infty}\frac{P(\phi^{\theta}, k)}{\sqrt k}= \theta. \eqno(*)
$$
For $\theta_1, \theta_2 \in (0, 1)$ and $\theta_1 \neq \theta_2$, we have 
$\phi^{{\theta}_1}\neq \phi^{{\theta}_2}$ from $(*)$. Hence $S$ is an 
uncountable subset of $F$.

From the construction of $\phi^{\theta}$, only the coordinates of 
type $c_{k^2}^{\theta}$ 
may take a non-zero value. Therefore, for $ \theta_1, \theta_2 \in (0, 1),
\theta_1 \neq \theta_2,$ from $(*)$, there exists an infinite number of positive
integers $k_n, n= 1, 2, \cdots,$ such that $c_{k_n^2}^{{\theta}_1}\neq 
c_{k_n^2}
^{{\theta}_2}$.

So $\forall r>0,$
\begin{eqnarray} 
p_r(B^{k_n^2}(\phi^{{\theta}_1}-\phi^{{\theta}_2})) & =  &
(\sum_{m=0}^{\infty}|c_{m+k_n^2}^{{\theta}_1}
-c_{m+k_n^2}^{{\theta}_2}|^2(m+1)(m+2)\cdots(m+k_n^2)(m+1)^r)^{1/2} 
		\nonumber  \\ 
& \geq & |c_{k_n^2}^{{\theta}_1}-c_{k_n^2}^{{\theta}_2}|^2k_n^2!= 1>0, \\
\liminf\limits_{n\rightarrow \infty}p_r(B^{k_n^2}
(\phi^{{\theta}_1}-\phi^{{\theta}_2})) & \geq & 1>0.
\end{eqnarray}
Therefore, we obtain
\begin{eqnarray} 
\limsup\limits_{k\rightarrow \infty} \rho(B^k(\phi^{{\theta}_1}), 
B^k(\phi^{{\theta}_2}))>0, 
\ \ \ \forall \theta_1\neq \theta_2.
\end{eqnarray} 
This proves condition (i) in the chaos definition of Li-Yorke.

Moreover,
$\forall k \geq 1,$ when $k^2+1 \leq l \leq (k+1)^2-1, c_l^{{\theta}_1} = 
c_l^{{\theta}_2} = 0, 
\forall \theta_1, \theta_2 \in (0, 1)$. So we have
 
\begin{eqnarray} 
p_r^2(B^{k^2+1}(\phi^{{\theta}_1}-\phi^{{\theta}_2}))
& = & \sum_{m=0}^{\infty}|c_{m+k^2+1}^{{\theta}_1}
-c_{m+k^2+1}^{{\theta}_2}|^2(m+1)(m+2)\cdots(m+k^2+1)(m+1)^r
					\nonumber \\ 
& \leq & \sum_{N=k+1}^{\infty}|c_{N^2}^{{\theta}_1}-
c_{N^2}^{{\theta}_2}|^2(N^2-k^2)(N^2-k^2+1)
\cdots N^2\cdot(N^2-k^2)^r 		\nonumber \\ 
& \leq & \sum_{N=k+1}^{\infty}\frac{(N^2-k^2)^r}{(N^2-k^2-1)!}
					\nonumber \\
& \leq & \sum_{N=k+1}^{\infty}\frac{(N^2-k^2)^r}{2^{N^2-k^2-2}} 
					\nonumber \\
& = & \sum_{m=1}^{\infty}{4(m^2+2mk)^r}/{2^{m^2+2mk}}
\end{eqnarray}
Thus, $\forall r>0,$ 
$$
0\leq \lim\limits_{k\rightarrow \infty}p_r(B^{k^2+1}(\phi^{{\theta}_1}-
\phi^{{\theta}_2}))
\leq \lim\limits_{k\rightarrow \infty}(\sum_{m=1}^{\infty}\frac{4(m^2+2mk)^r}
{2^{m^2+2mk}})^{1/2}= 0,
$$
which implies that
\begin{eqnarray} 
\liminf\limits_{k\rightarrow \infty} \rho(B^k(\phi^{{\theta}_1}), 
B^k(\phi^{{\theta}_2}))= 0.
\end{eqnarray}
This proves condition (ii) of the Li-Yorke chaos definition.
Therefore, $S$ is a chaotic set for $B$, and $(F, B) $ is  a chaotic system 
in the sense of Li-Yorke.
  
\section{Remarks}

Recently, Godefroy and Shapiro (\cite{Godefroy_Shapiro}) have shown 
that the weighted backward shift operator in a separable Hilbert space
$H$, with a complete orthonormal basis $\{ \phi_n \}$,
$$
\begin{array}{c}
b: H\rightarrow H,\\ 
b\phi_n = \mu \phi_{n-1},
\end{array}
$$  
is chaotic in the sense of Devaney, whenever $|\mu|>1$.
 
Following the method given in the last section, we can also discuss the 
chaos of the 
operator $b$ in the sense of Li-Yorke. When $|\mu| >1 $ , the chaotic 
set $S$ of $b$ can be constructed as follows:
$$
S = \{ \phi^{\theta} = \sum_{n=0}^{\infty}c_n^{\theta}\phi_n : \theta \in 
(0, 1) \},
$$
where
$$
\left\{\begin{array}{l}
c_0^{\theta} = 0  \\
c_n^{\theta} = \left\{\begin{array}{cl} 1/{\mu^n} & \mbox{ if $n=k^2,[k\theta]-
[(k-1)\theta]=1, k\geq 1 $}  \\
0  & \mbox{otherwise .}
\end{array}\right.  
\end{array}\right.
$$
Hence $b$ is also chaotic in the sense of Li-Yorke whenever $|\mu| >1 $.

When $\mu = 1, b$ is similar to the left shift map $\sigma$ in 
symbolic dynamics. However,  when $|\mu| \leq 1$, 
the global attractor 
of $b$ is the one-point set  containing only the  
zero  vector;  thus $b$ is  not chaotic in the sense of Li-Yorke or Devaney.

On a different phase space, i.e., on the Fr\'{e}chet space 
$\Sigma(X) = \{(x_0, x_1, \cdots, x_k, \cdots): x_k\in X, k\geq 0 \}$, 
where $X$ is a non-trivial Fr\'{e}chet space, and $\Sigma(X)$ 
is equipped with product topology, Fu and Duan (\cite{Fu_Duan}) have shown 
that $\sigma$ is chaotic on $\Sigma(X)$ in the senses of both Li-Yorke 
and Devaney (or Wiggins). 
 
\medskip
 
{\bf Acknowledgement}  
	We would like to thank Drs. Vince Ervin and
John Lawson, both of Clemson University, for useful discussions.

\end{document}